\documentclass[times,9pt,twocolumn]{article}
\usepackage{latex8}
\usepackage{times}
\usepackage{amssymb}
\usepackage{xspace}
\usepackage{bm}
\usepackage{cite}
\usepackage{hero}
\usepackage{amsmath}
\usepackage{graphicx}

\include{def_macros}
\def\omegarf{\omega_{{\mathit{rf}}}}

\def\Tskip{T_{{\mathit{skip}}}}

\usepackage{spconf}

\begin{document}
%Changed Title 
\title{\textbf{Baseband Detection of Bistatic Electron Spin Signals in
Magnetic Resonance Force Microscopy (MRFM) } }

\name {Chun-yu Yip$^*$, Alfred O. Hero$^*$, Daniel Rugar$^\dag$, Jeffrey A. Fessler$^*$ \thanks{Research
partially supported by DARPA Mosaic program under ARO contract
DAAD19-02-C-0055. }%
\thanks{Corresponding author: C.Y. Yip (chunyuy@umich.edu)}}
\address{$^{*}$  University of Michigan, Ann Arbor, MI\\
  $^{\dag}$IBM Research Division, Almaden Research Center, San Jose, CA}

\maketitle
\thispagestyle{empty}

\bc
\bf Abstract
\ec

\textit{In single spin Magnetic Resonance Force Microscopy 
(MRFM), the objective is to detect the presence of an electron 
(or nuclear) spin in a sample volume by measuring spin-induced 
attonewton forces using a micromachined cantilever. In the OSCAR 
method of single spin MRFM, the spins are manipulated by an 
external rf field to produce small periodic deviations in the 
resonant frequency of the cantilever. These deviations can be 
detected by frequency demodulation followed by conventional 
amplitude or energy detection. In this paper, we present an 
alternative to these detection methods, based on optimal 
detection theory and Gibbs sampling.  On the basis of 
simulations, we show that our detector outperforms the 
conventional amplitude and energy detectors for realistic MRFM 
operating conditions. For example, to achieve a 10\% false alarm 
rate and an 80\% correct detection rate our detector has an 8 dB 
SNR advantage as compared with the conventional  amplitude or 
energy detectors. Furthermore, at these detection rates it comes 
within 4 dB of the omniscient matched-filter lower bound. }

\section{Introduction}

Magnetic Resonance Force Microscopy (MRFM) is a recently 
developed technique with which physicists can potentially push 
the limits of force detection to the single electron spin level, 
with sub-angstrom spatial resolution 
\cite{SIDLES:NONDESTRUCT,SIDLES:OSCILL,RUGAR:FORCEDET1}. The 
experiment involves detection of perturbations of a thin 
micrometer-scale cantilever whose tip incorporates a submicron 
ferromagnet.  Any spinning electrons in the sample will act as 
magnetic dipoles, exerting perturbing forces that can be measured 
from cantilever displacements. There have been several successful 
experimental demonstrations of MRFM for imaging micron-size 
ensembles of spins. For example, three-dimensional imaging with 
micrometer spatial resolution has been achieved 
\cite{ZUGER:THREED}. Furthermore, forces as small as $8 \times 
10^{-19}$ N have been detected using MRFM 
\cite{Mamin&Rugar:AP01}.  However, despite several advances, 
detection of an isolated single electron spin has not yet been 
accomplished. Progress towards this goal will require advances in 
physical measurements and advances in signal processing of these 
measurements.

Recently a MRFM method known as OScillating Cantilever-driven Adiabatic
Reversals (OSCAR)~\cite{STIPE:RELAX} has been proposed to detect single
spins. This method, explained below, uses a modulated external radio
frequency (rf) magnetic field to manipulate the electron spins, in order to produce
periodic forces on the cantilever that can be detected as small frequency
shifts.  Detection of these frequency shifts identifies the presence of the
electron spin. 
If successful, single electron spin detection would be an important step
towards the long-term goal of three-dimensional imaging of subsurface atomic
structure \cite{SIDLES:MRFM}.

Unfortunately, accurate single-spin detection in OSCAR is hampered by several
factors. The spin-induced frequency shift signal is extremely weak as a spin
induces a frequency shift of only one part in $10^4$.
Thus long integration times are required to detect such a signal. 
However, spin relaxation and decoherence significantly reduce the 
usable integration time, especially at room temperature.  This 
makes the use of cryogenics (cooling the experimental apparatus 
down to a fraction of a degree Kelvin) necessary to reduce these 
effects. However, in this regime the measurements are prone to 
thermal noise from various sources. Noise and decoherence effects 
must be taken into account by the detection algorithm to 
achieve the most accurate and reliable single spin detection.  
Very simple detectors are the baseband ``amplitude detector'' and 
``energy detector'' which operate on a frequency demodulated 
version of the cantilever position signal.  Such detection 
schemes are widely used in MRFM, NMR spectroscopy, MRI, and other 
applications.

In this paper, we present a new approach to baseband detection in 
OSCAR experiments. The detector is based on a random telegraph 
model for the baseband signal incorporating Poisson-distributed 
random spin reversals, random initial spin polarity, and Additive 
White Gaussian Noise (AWGN). To accurately decide 
between the hypotheses of spin absence and presence, we propose a 
hybrid detection scheme which combines optimal \emph{Bayes} and 
\emph{General Likelihood Ratio} (GLR) detection principles 
implemented with Gibbs sampling.  Simulations show that our 
proposed detector can significantly outperform the conventional 
baseband amplitude and energy detector for realistic 
post-demodulation signal-to-noise ratios (SNR).

The outline of the paper is as follows. After briefly reviewing the OSCAR
experiment in Sec.~2, we describe the proposed signal detection scheme in
Sec.~3, and present results of numerical simulations in Sec.~4.

\section{Description of Experiment}

\begin{figure}[htb]
\begin{center}
%\begin{minipage}[b]{1.0\linewidth}
%\centering
\includegraphics[width=3.25in]{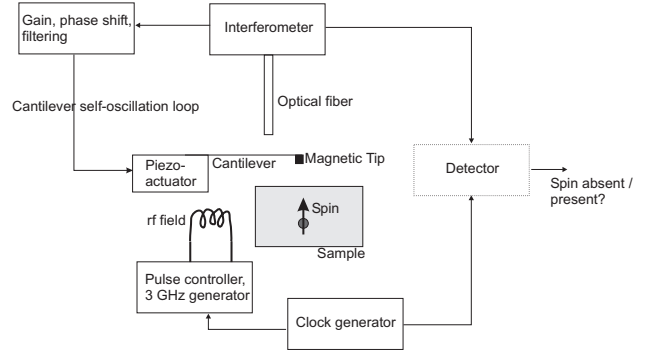} 
%\end{minipage}
%
\caption{\small \textit{Schematic of the MRFM experiment.}}
\label{fig:setup}
\end{center}
\end{figure}

Fig.~\ref{fig:setup} is a schematic description of the OSCAR 
experiment. In OSCAR, a submicron ferromagnet is placed at the 
tip of a cantilever which sits at a distance of approximately 
$50$ nanometers above a sample. In the presence of an applied rf 
field, electrons in the sample undergo magnetic resonance if the 
rf field frequency matches the Larmor frequency. Since the Larmor 
frequency is proportional to the field from the magnetic tip, and 
because the tip field falls off rapidly with distance, only those 
spins that are within a thin ``resonant slice''  just the right 
distance from the tip will satisfy the condition for magnetic 
resonance and contribute to the signal.

If the cantilever is forced into mechanical oscillation by positive feedback, the tip oscillation induces small shifts in the Larmor frequencies of the
spins. Specifically, the tip motion gives rise to an oscillating magnetic
field which sweeps the Larmor frequency of the spins in the resonance slice
back and forth through resonance. This causes the spin to reverse polarity
synchronously with the cantilever motion, and in return, the spin reversals
affect the cantilever motion by changing the effective stiffness of the
cantilever. When an electron spin is present the spin-cantilever interaction
can be detected by measuring small shifts in the period of cantilever
oscillation using laser interferometric cantilever position sensing. For more
details about OSCAR, see \cite{SIDLES:MRFM,RUGAR:FORCEDET2,WAGO:FD}.

\begin{figure}[!htb]
\begin{center}
%\begin{minipage}[b]{1.0\linewidth}
%\centering
\includegraphics[width=2.5in]{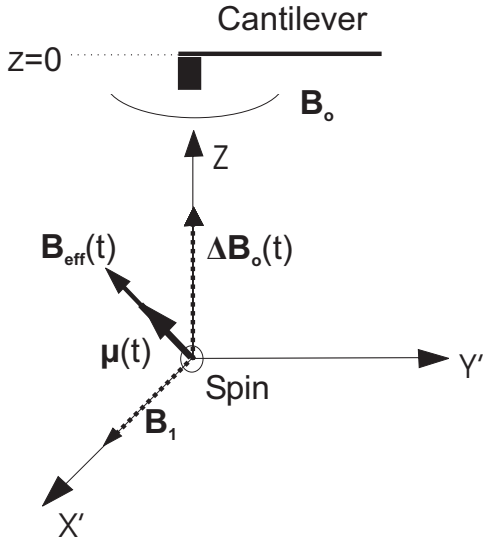} 
%\end{minipage}
%
\caption{\small \textit{In the coordinate system rotating at $\omegarf$, the
off-resonance field $\Delta \bB_{o}$, and therefore the effective field
$\mathbf{B}_{\mathit{eff}}(t)$, vary with time. Under the spin lock
assumption, the spin always follows $\pm\mathbf{B}_{\mathit{eff}}(t)$.}}
\label{fig:coord}
\end{center}
\end{figure}

A classical (non-quantum) electro-mechanical description of the
spin-cantilever interactions can be developed in the framework adopted by
Berman {\it et al} \cite{BERMAN:OSCAR} and Rugar {\it et al}
\cite{Rugar:Memo02}. We briefly review this framework here. Consider a spin
in a rotating frame which rotates at the frequency of the applied rf magnetic
field, $\mathbf{B}_{1}$ (Fig.~\ref{fig:coord}). The effective magnetic field
$\mathbf{B}_{\mathit{eff}}(t)$ in this frame is given by
\begin{eqnarray}
\mathbf{B}_{\mathit{eff}}(t) & = & B_{1}\mbox{\boldmath $\hat{i}$} + \Delta B_{o}(t) \mbox{\boldmath $\hat{k}$}, 
\end{eqnarray}
where \mbox{\boldmath $\hat{i}$} and \mbox{\boldmath $\hat{k}$} are unit
vectors in the $x$ and $z$ directions in the rotating frame, $B_{1}$ is the
amplitude of the rf magnetic field, $B_{o}(t)$ is the magnetic field caused by
the magnetic tip of the cantilever, and $\Delta B_{o}(t) = B_{o}(t) -
\omegarf/\gamma$ is the off-resonance field magnitude ($\gamma$ is the
gyromagnetic ratio). Spins are in resonance on the spherical shell (resonant
slice) defined by those spatial locations for which $\omegarf$ matches the
Larmor frequency $\gamma B_{o}(t)$.

If $\Delta B_{o}(t)$ varies sufficiently slowly such that the adiabatic
criterion
\begin{eqnarray}
\frac{d\Delta B_{o}(t)}{dt} & \ll & \gamma B_{1}^{2}
\end{eqnarray}
is met, the spin can be assumed to remain aligned or anti-aligned
with the vector $\mathbf{B}_{\mathit{eff}}(t)$. This is the \emph{spin-lock}
condition. Let the vertical position of the cantilever tip be denoted by $z$
where $z=0$ denotes its rest position. Under the influence of the
external rf field $\bB_1(t)$, electron-spin forces, and random (thermal)
force noise $F_n(t)$, the motion of the cantilever tip obeys the simple
harmonic oscillator equation:
\begin{eqnarray}
m\ddot{z}(t) + \Gamma\dot{z}(t) + kz(t) & = & \frac{|\mbox{\boldmath $\mu$}| G^2 z(t)}{\sqrt{G^2 z(t)^2 + B_{1}^2}} + F_{n}(t), \nonumber\\ 
\label{eq:shoe}
\end{eqnarray}
where $m$ is the cantilever's effective mass, $k$ is the cantilever spring
constant, $\Gamma$ is the friction coefficient characterizing cantilever
energy dissipation, $|\mbox{\boldmath $\mu$}|$ is the magnitude of the spin
magnetic moment, $G = \partial B_{oz}/\partial z$ is the z-direction field
gradient at the spin location. The natural mechanical resonance frequency of
the cantilever is given by $\omega_{o} = \sqrt{k/m}$, and $\Gamma$ can be
related to the cantilever quality factor, $Q$, via $\Gamma =
k/(\omega_{o}Q)$.
 
Under the small tip displacement approximation $|Gz|\ll B_{1}$, we obtain
\begin{eqnarray}
m\ddot{z}(t) + \Gamma\dot{z}(t) + (k + \Delta k) z(t) & \approx & F_{n}(t), 
\end{eqnarray}
where $\Delta k = -|\mbox{\boldmath $\mu$}| G^2/ B_{1}$. This shift in spring
constant results in a shift $\Delta \omega_o$ of the cantilever resonant
frequency:
\begin{eqnarray}
\label{eq:shift}
\Delta \omega_{o} & \approx & - \frac{1}{2} \omega_{o} \frac{|\mbox{\boldmath
    $\mu$}| G^2}{k B_{1} }. 
\end{eqnarray}
In a version of the protocol called ``Interrupted OSCAR,'' the $\bB_{1}$
field is turned off every $\Tskip$ seconds over a half cycle duration
($\pi/\omega_o$) to cause periodic transitions between the \emph{spin-lock}
and \emph{anti-spin-lock} spin states (see middle panel of
Fig.~\ref{fig:signals}). In the spin-lock state the spin aligns with the
field $\mathbf{B}_{\mathit{eff}}(t)$ and in the anti-lock state the spin
aligns with $-\mathbf{B}_{\mathit{eff}}(t)$. Therefore, the frequency shift
$\Delta \omega_o$ of the cantilever alternates between the two values
$\pm\frac{1}{2}\omega_{o}(|\mbox{\boldmath $\mu$}|) G^2/(k B_{1})$ with period
$\Tskip$.  In the absence of noise ($F_n(t)=0$ in (\ref{eq:shoe})) the
cantilever motion can be expressed as the frequency modulated (FM) signal:
\begin{eqnarray}
\label{eq:zsignal}
z(t) &=& A\cos\left(\omega_{o}t+\int_{0}^{t} \bar{s}(t') dt' + \theta\right).
\end{eqnarray}
Here $A$ is the cantilever oscillation amplitude, $\theta$ is a 
random phase, and $\bar{s}$ is equal to $0$ if no spin coupling 
occurs, while it is equal to a periodic square wave of period 
$2\Tskip$ and of amplitude $|\Delta\omega_o|$ if spin coupling 
occurs.  Thus, in this ideal noiseless case, the presence of spin 
coupling can be perfectly detected either by detecting a spectral 
peak near $|\Delta\omega_o|$ radians in the periodogram or by 
frequency demodulation of $z$ to baseband (incorporating 
subtraction of the known center frequency $\omega_o$) followed by 
amplitude detection, energy detection, or other algorithm, as 
discussed below.  As baseband and narrowband are equivalent 
representations we focus on the baseband method here.  These 
methods correlate the baseband signal against the known square 
wave signal derived from $\bB_1$. The resultant signal, which we 
call $y(t)$, forms the statistic which is used for spin 
detection, as illustrated in Fig.~\ref{fig:baseband}.

Unfortunately, in a practical (non-ideal) experiment the 
cantilever tip-displacement signal $z(t)$ is degraded by several 
factors which reduce correlation peak detection accuracy.  One 
factor is the presence of laser interferometric measurement 
noise. This adds a noise floor to the demodulated square wave 
signal $\bar{s}(t)$.  Another factor is spin relaxation which 
over a period of time causes the spin to go out of alignment with 
the effective field $\mathbf{B}_{\mathit{eff}}$. While several 
models for single-spin relaxation have been proposed 
\cite{Rugar:Memo02,BERMAN:OSCAR}, a full understanding of the 
physics of single-spin relaxation interactions with cantilevers 
remains open.  One model, adopted here, is that the single spins 
maintain spin lock or anti-lock states but spontaneously and 
asychronously change polarity during the course of measurement at 
some rate $\lambda$ flips/second.  In the sequel we develop an 
optimal single-spin detection approach under a random Poisson 
model for these polarity flips.

\section{Signal Detection in Noise}
\begin{figure}[!htb]
\begin{center}
%\begin{minipage}[b]{1.0\linewidth}
%\centering
\includegraphics[width=3in]{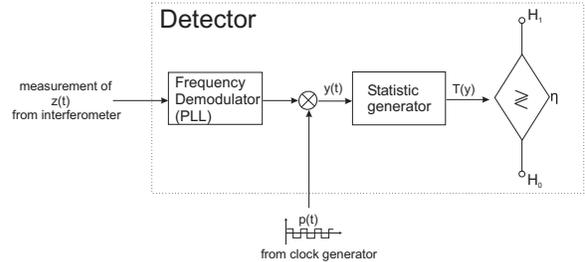}
%\end{minipage}
%
\caption{\small \textit{Baseband detector frequency demodulates the
interferometric signal, correlates the output against a square wave $p(t)$
whose transitions are synchronous with the turn-off times of the rf field
$\bB_1(t)$, and generates a test statistic, e.g.~accumulated squared
frequency deviations, for detecting presence of a spin.}}
\label{fig:baseband}
\end{center}
\end{figure}

The signal detectors we will consider operate on the baseband output signal
$y(t)$ of the frequency demodulator, e.g.~a Phase-Lock Loop (PLL), followed
by multiplication by a square wave reference $p(t) \in \{\pm1\}$ of period
$2\Tskip$, whose transitions are synchronous with the (known) rf turn-off
times (see Fig. \ref{fig:baseband}).

\begin{figure}[!htb]
\begin{center}
%\begin{minipage}[b]{1.0\linewidth}
%\centering
\includegraphics[width=3.0in]{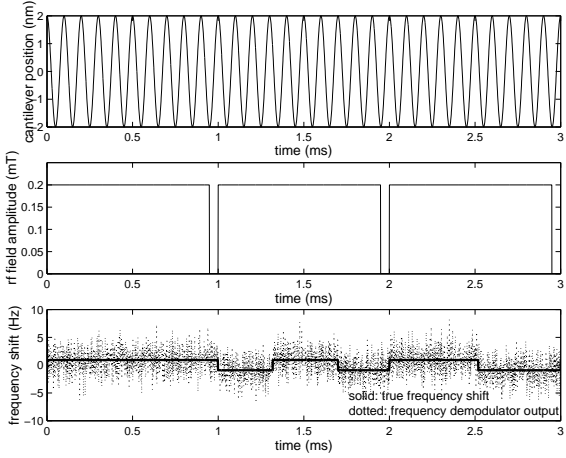}
 
%\end{minipage}
%
\caption{\small \textit{Top: Sample cantilever position signal, $z(t)$, at 10
kHz. Middle: Sample rf magnetic field magnitude, $B_{1}$, has synchronous
half-cycle skips at 1 ms, 2 ms, and 3 ms. Bottom: In the presence of a single
spin, $\bar{s}$ in Eq. \ref{eq:zsignal} has both deterministic transitions
due to the rf skips at 1 ms, 2 ms and 3 ms, and random ones due to spin
relaxation. The random transitions, $\{\tau_{i}\}$, occur as a Poisson
process. The initial polarity is $\phi=1$ for this example. The noisy signal
at bottom is $\bar{s}$ with AWGN contamination.}}  
\label{fig:signals}
\end{center}
\end{figure}

We model the baseband output $y(t)$ of the frequency demodulator and
correlator as a random telegraph plus additive Gaussian white noise (see
lower panel of Fig.~\ref{fig:signals}).  Let $[0,T]$ be the total measurement
time period over which the correlator integrates the measurements, and let
$\{\tau_{i}\}, i =1..N$, be the time instants within this period at which
spin reversals occur. We assume $\{\tau_{i}\}$ are the arrival times of a
Poisson process with intensity $\lambda$. Consequently $N$ is a Poisson
random variable with rate $\lambda T$ \cite{DAVEN:PROB}.  Thus,
$y(t)=s(t)+v(t)$ where $v(t)$ is AWGN with variance = $\sigma_{v}^{2}$, and
$s(t)$ is a random telegraph signal containing only the random transitions:
\begin{eqnarray}
s(t) & = &
\phi|\Delta\omega_{o}|\sum_{i=0}^{N}(-1)^{i}g(\frac{t-\tau_{i}}{\tau_{i+1}-\tau_{i}}), 
\label{eq:strt}
\end{eqnarray}
where $\phi$ is a random variable that takes on $\pm1$ with equal
probability, representing a random initial spin polarity, $\tau_{0}=0$,
$\tau_{N+1}=T$, and $g(t)$ is the standard rectangle function: $g(t)=1$ for
$t\in[0,1]$ and $g(t)=0$ otherwise.  If there are no random spin flips in the
time period $[0,T]$, then $s(t)=\phi|\Delta\omega_{o}|$ is constant
over time, which we obtain in (\ref{eq:strt}) by using the convention that
when $N=0$, $\tau_o=0$ and $\tau_1=\infty$.

The baseband spin detection problem is to design a test between the two
hypotheses:
\begin{eqnarray}
H_{0} \textrm{ (spin absent):  }  &  y(t)& = v(t) \nonumber\\
H_{1} \textrm{ (spin present):  } &  y(t)& = s(t) + v(t)
\label{eq:nine}
\end{eqnarray}
for $t\in[0,T]$.% (Fig. \ref{fig:signals}). 

Conditioned on the random parameters $\{\tau_{i}\},N,\phi$, the signal $s(t)$
is deterministic and known. Under this conditioning the \emph{optimal}
detection structure would be the simple matched filter \cite{POOR:INTRO}
\begin{eqnarray}
\label{eq:MF}
\frac{1}{T}\int_{0}^{T}y(t')s(t',\phi,\boldsymbol{\tau},N)dt' 
 \comp \eta
%\gtrless_{H_{0}}^{H_{1}} 
\end{eqnarray}
where $s(t;\phi,\boldsymbol{\tau},N)$ is a synthesized random telegraph
signal of the form (\ref{eq:strt}) parametrized by $|\Delta\omega_o|$
(assumed known), $\phi,\boldsymbol{\tau}$ and $N$. The value $\eta$ is a
threshold that can either be set to satisfy a \emph{probability of false
alarm} ($P_{F}$) contraint $P_{F} \leq \alpha$, $\alpha \in [0,1]$, or as a function of
the prior probabilities $a\ln [P(H_0)/P(H_1)] +b$ where $a,b$ are known
constants.
%a=\sigma_{v}^{2} +
%\int_{0}^{T}[s^+(t',\boldsymbol{\tau},N)]^2 dt'/2$. 
%b=2^{-1}\sigma_{v}^{2}\ln(2\pi\sigma_v^2) 
In the former case the detector is called the {\it most powerful} (MP) test
of level $\alpha$, which has maximum \emph{probability of detection}
($P_{D}$), while in the latter case the detector is called the minPe detector
as it achieves minimum average probability of decision error (minPe).

As the values of the random parameters are always unknown, we call the
detector (\ref{eq:MF}) the {\it omniscient matched filter}, which is
unimplementable. However, as the omniscient matched filter is optimal for
known parameter values it establishes a useful upper bound on performance.

Perhaps the simplest baseband detection scheme, and the most widespread in
MRFM applications, is the \emph{amplitude detector} which acts as if there
were no random flips and declares a spin present if the magnitude of the
average amplitude of the correlator output exceeds a threshold
%(also known as a \emph{radiometer}) 
%\cite{POOR:INTRO}
%
\begin{eqnarray}
\label{eq:AD}
\left|\frac{1}{T}\int_{0}^{T}y(t') dt'\right| \comp
%\gtrless_{H_{0}}^{H_{1}} 
\eta
\end{eqnarray}
where $\eta$ is a threshold set to give the desired $P_{F}$.  Improved
 performance can be obtained by explicitly accounting for the equally likely
 initial polarity and assuming AWGN to derive the minPe detector. The
 amplitude detector (\ref{eq:AD}) is the minPe detector under the assumption
 that $y(t)$ is a random polarity constant embedded in AWGN.  This is a valid
 assumption when there are no random spin flips over the integration period 
 $[0,T]$.

When there are random spin flips over $[0,T]$ due to spin relaxation and decoherence, the
performance of the amplitude detector suffers. Indeed, as the
number of random flips increases the average amplitude of $y(t)$ converges to
zero. As the energy of $s(t)$ is independent of number of transitions, transition times, and initial polarity, it is natural to propose an
\emph{energy detector}
%(also known as a \emph{radiometer}) 
\cite{POOR:INTRO}
\begin{eqnarray}
\label{eq:ED}
\int_{0}^{T}[y(t')]^2 dt' \comp
%\gtrless_{H_{0}}^{H_{1}} 
\eta
\end{eqnarray}
where $\eta$ is a threshold set to give the desired $P_{F}$. It can be
shown that the energy detector is a minPe test for the case that $v(t)$ is
additive white gaussian noise, $s(t)= \Delta \omega_o \cos(2 \pi t /\Tskip
+ \theta)$, and $\theta$ is uniformly distributed over $[0, 2\pi]$
\cite{VANTREES:DETECTION}. It can also be
shown that the energy detector is the minPe test under a Gaussian
approximation to the random telegraph process in the limit of high SNR
\cite{Hero&etal:OSCAR03}. Note that in our case the detection performance of the energy test is independent of the average flip rate $\lambda$.

As we will show in the sequel, the performance of the amplitude and energy
detectors can be far from the optimal performance achieved by the omniscient
matched filter detector.

\subsection {The Hybrid Bayes/GLR Detector}
The minPe detector for a signal with random parameters is a Bayes likelihood
ratio test that averages an omniscient likelihood ratio test statistic over
all random parameters \cite{VANTREES:DETECTION}:
\begin{eqnarray}
\label{eq:optimal}
&&\log \Lambda(y)
\\ & & =\log \frac{
E_{\boldsymbol{\tau},N}\left[ E_{\phi} \left[ f \left( y;
\boldsymbol{\tau},N, \phi | H_{1}\right) \right] \right]} 
{ f(y | H_{0} ) } \comp 
%\gtrless_{H_{0}}^{H_{1}} 
\eta \nonumber .
\end{eqnarray} 
As above $\eta$ is a threshold selected to achieve a desired level $\alpha$
of $P_{F}$. The function $f$ is the joint p.d.f of $\{y(t)\}_{t \in [0,T]}$
parametrized by the random parameters $\boldsymbol{\tau},N, \phi$, and
$E_{\mathbf{x}}[\cdot|A]$ denotes conditional expectation over random
variables $\mathbf{x}$ given event $A$.
%Note that the logorithm only affects the threshold, but not the
%decision regions.

While the expectation over $\phi$ in (\ref{eq:optimal}) is simple 
to evaluate, the expectation over $\{\{\tau_i\}, N\}$ is very 
difficult since the integration region is of very high (infinite) 
dimension. An alternative to performing this second 
expectation is to invoke the Generalized Likelihood Ratio (GLR) 
principle. The GLR consists of replacing the unknown parameters 
in (\ref{eq:optimal}) by \emph{Maximum Likelihood} (ML) estimates.
\begin{eqnarray}
\label{eq:GLRT}
\lefteqn{\log \Lambda(y)} \\ & & =\log \frac{
\max_{\boldsymbol{\tau},N} \left\{E_{\phi} \left[ f
\left( y;\boldsymbol{\tau},N,\phi | H_{1}\right) \right] \right\}}{f
(y | H_{0})} \comp 
%\gtrless_{H_{0}}^{H_{1}} 
\eta \nonumber ,
\end{eqnarray} 
where, again, $\eta$ is a threshold chosen for a desired $P_{F}$. 
Note that in (\ref{eq:GLRT}) we have averaged over $\phi$ while 
we have maximized over $\{\{\tau_{i}\},N\}$, leading to what we 
call a \emph{hybrid Bayes/GLR} test. 

As $y(t)$ is a conditionally Gaussian random process given $\{{\tau_{i}}\}$
and $N$, the log-likelihood function in (\ref{eq:GLRT}) can be simplified by
invoking the Cameron-Martin formula \cite{POOR:PARA}:
\begin{eqnarray}
%\lefteqn{
\log \Lambda(y) & \!\!\!\!=\!\!\!\! & \max_{\boldsymbol{\tau},N} \bigg\{ \log \cosh \bigg[
\frac{1}{\sigma_{v}^{2}}
 \int_{0}^{T} \!y(t)s^{+}(t;\boldsymbol{\tau},N)dt
\bigg] \bigg\} \nonumber  \\ & \!\!\!\!-\!\!\!\!
&\frac{1}{\sigma_{v}^{2}}\int_{0}^{T}\!(s^{+}(t;\boldsymbol{\tau},N))^{2} dt
 \label{eq:CM}
\end{eqnarray}
where $s^{+}(t;\boldsymbol{\tau},N)$ is the synthesized telegraph 
wave (\ref{eq:strt}) having initial polarity $\phi=1$ and 
parametrized by $\boldsymbol{\tau}$ and $N$. It is well known 
that for a sufficiently large integration time $T$ the minPe and 
GLR tests are identical (see for example \cite{LeCam:86}). Thus 
we can assert that the hybrid Bayes/GLR test (\ref{eq:GLRT}) is 
an asymptotically optimal test.

\subsection{Solution via Gibbs Sampling}
The maximization in (\ref{eq:CM}) by exhaustive search over the uncountably
infinite dimensional space of possible parameter values, $\{\{\tau_{i}\},N\}$, is
impractical.  An alternative, which is guaranteed to converge to the
maximizing solution, is to more efficiently search over the space by
Gibbs Sampling \cite{ROBERT:MCMC,BREMAUD:99}.  As we know the Poisson
intensity $\lambda$, we can generate samples $\{\{\tau_{i}\},N\}$ from the
\emph{prior} Poisson distribution so as to maximize the log-likelihood
function. As these samples are more likely (on the average) to mimic the
actual behavior of the parameters, we obtain a reduction in
search complexity.

The general description of the Gibbs sampler is as follows. Supposed there is
a random vector variable $\mathbf{X}=[x_{1},x_{2},\ldots,x_{p}]^{T}$ having
density function $f_{\mathbf{X}}$ from which we want to sample. Suppose also
that we can simulate the i-th element of $\mathbf{X}$ given samples (already
simulated) of the other elements:
\begin{eqnarray}
 & X_{i} | x_{1},x_{2}, \ldots, x_{i-1}, x_{i+1},\ldots, x_{p} & \nonumber \\
& \sim f_{i}(x_{i}|x_{1},x_{2}, \ldots, x_{i-1}, x_{i+1},\ldots, x_{p}) & \textrm{ for } i = 1..p \nonumber \\
\end{eqnarray}
Then a Markov sequence, $\mathbf{x}^{(t)} = [x_{1}^{(t)}, \ldots,
x_{p}^{(t)}]^{T}$, can be simulated by the recursion
\begin{eqnarray}
X_{1}^{(t+1)} &  \sim & f_{1}(x_{1}|x_{2}^{(t)}, \ldots, x_{p}^{(t)}),  \nonumber\\
X_{2}^{(t+1)} &  \sim & f_{2}(x_{2}|x_{1}^{(t+1)}, x_{3}^{(t)}, \ldots, x_{p}^{(t)}), \nonumber\\
\!\!\!\!\!\!\!\!\!\!\!\!\!\!\!\!\!\!\!\! \vdots & & \nonumber\\
X_{p}^{(t+1)} & \sim  & f_{p}(x_{p}|x_{1}^{(t+1)}, x_{2}^{(t+1)}, \ldots, x_{p-1}^{(t+1)}). \nonumber \\
\label{eq:Gibbs1}
\end{eqnarray}
After a certain amount of burn-in time $T_{b}$, 
$\mathbf{X}^{(t)}, t>T_{b},$ will have stationary distribution 
$f_{\mathbf{X}}$. In our case, since the arrival times 
$\{\tau_{i}\}$ are generated from a Poisson process, the 
conditional distributions (\ref{eq:Gibbs1}) are easy to sample 
from, because they are conditionally uniform.

\section{Simulation Methods and Results}
The objective of our first three simulations was to compare the detection
performance of the matched filter, the amplitude detector, the energy
detector, and the Bayes/GLR detector on the basis of Receiver Operating
Characteristic (ROC) curves, which are obtained by empirically generating the
pairs $(P_{F},P_{D})$ for each detector. In our simulations, the four
decision rules (\ref{eq:MF}), (\ref{eq:AD}), (\ref{eq:ED}), and
(\ref{eq:GLRT}) were used to generate the ROC curves in the Matlab 6.1
environment. Based on the Monte Carlo methodology \cite{ROBERT:MCMC}, we
generated samples $\{y_{d}^{(i)}(n)\}$, $y_{d}(n)=y(nT_{s})$, under both
Hypothesis 0 and 1, where $T_{s}$ was the sampling period. The samples were
input to the detector being evaluated, and $P_D$ and $P_F$ were statistically
calculated. 500 detection trials were performed under each hypothesis.
% to obtain each pair of probability values $(P_F, P_D)$. 
For each ROC curve, the above process was repeated with a range of decision
threshold values $\eta$. This range of thresholds was chosen to
adequately sample the domain $P_{F}\in[0,1]$. 
%For the range of experimental
%parameters investigated the cosh amplitude detector (\ref{eq:ADp}) and the
%ordinary amplitude detector (\ref{eq:AD}) had virtually identical
%performance so only the latter detector's performance is compared below.

The simulation parameter values were chosen according to typical 
OSCAR experimental values.  The signal duration $T$ was $3$ s and 
the sampling period $T_s$ was $0.5$ ms. The rest of the  
parameters were set to:
%$B_{1}=$2 Gauss, $G=2\times10^{10}$ Gauss/m,
%$\omega_{o}=$10 kHz, and $|\mbox{\boldmath $\mu$}| = 9.28\times10^{-28}$
%J/Gauss. 
$k=1 \times 10^{-3}$ N/m, $\omega_{o}/(2\pi)=1 \times 10^4$ Hz, 
$B_1=0.2$ mT, $G=2 \times 10^6$ T/m, and $\mu= 9.3 \times 
10^{-24}$ J/T.  With these parameters the (noiseless) signal 
amplitude, $|s(t)|$, was $0.928$ Hz according to 
(\ref{eq:shift}).  Two values of $\lambda$, the average number of 
random flips per second, were evaluated. The detector noise was 
assumed AWGN and the noise variance was adjusted to investigate 
the effect of SNR, which is defined as $10 
log_{10}[(1/T)\int_0^T|s(t)|^{2} dt / \sigma_{v}^{2}]$.

\begin{figure}[!htb]
\begin{center}
%\begin{minipage}[b]{1.0\linewidth}
%\centering
%\includegraphics[width=3.0in]{SNR25beta1000withMEAN_G5000c.eps}
\includegraphics[width=3.0in]{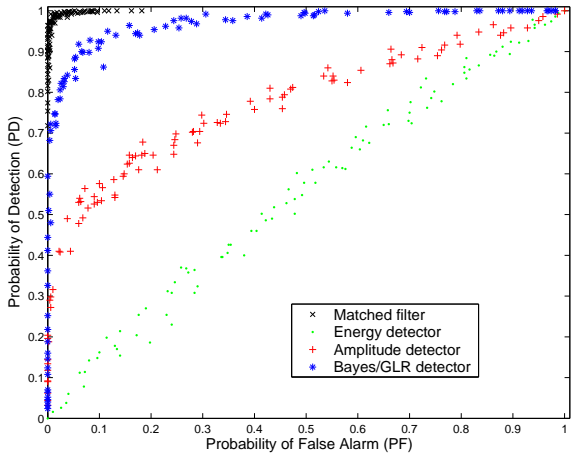} 
%\end{minipage}
%
\caption{\small \textit{Simulated Receiver Operating Characteristic (ROC)
curves for the matched filter, energy detector, amplitude detector, and
hybrid Bayes/GLR detector, at SNR = -25 dB and $\lambda=1$
event-per-second. Unlike the other detectors, the matched filter assumes
complete information on the parameter values and is not implementable.}}
\label{fig:ROC15_2}
\end{center}
\end{figure}

\begin{figure}[!htb]
\begin{center}
%\begin{minipage}[b]{1.0\linewidth}
%\centering
%\includegraphics[width=3.1in]{SNR20beta1000withMEAN_G5000c.eps} 
\includegraphics[width=3.0in]{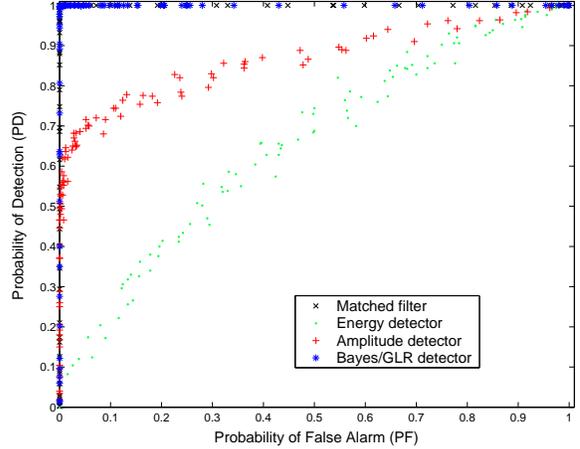}
%\end{minipage}
%
\caption{\small \textit{Simulated Receiver Operating Characteristic (ROC)
curves for the matched filter, energy detector, amplitude detector, and
hybrid Bayes/GLR detector, at SNR = -20 dB and $\lambda=1$
event-per-second. }}
\label{fig:ROC15_2p}
\end{center}
\end{figure}

\begin{figure}[!htb]
\begin{center}
%\begin{minipage}[b]{1.0\linewidth}
%\centering
%\includegraphics[width=3.0in]{SNR20beta100withMEAN_G5000c.eps} 
\includegraphics[width=3.0in]{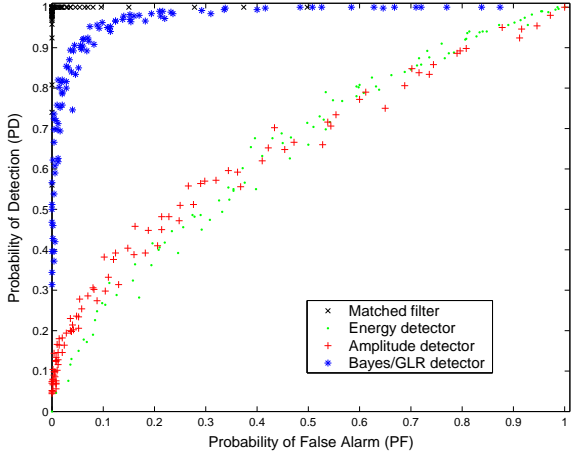}
%\end{minipage}
%
\caption{\small \textit{
Simulated Receiver Operating Characteristic (ROC)
curves for the matched filter, energy detector, amplitude detector, and
hybrid Bayes/GLR detector, at SNR = -20 dB and $\lambda=10$
events-per-second. }}
\label{fig:ROC175_2}
\end{center}
\end{figure}

\begin{figure}[!htb]
\begin{center}
%\begin{minipage}[b]{1.0\linewidth}
%\centering
%\includegraphics[width=3.0in]{powercurves.eps} 
\includegraphics[width=3.0in]{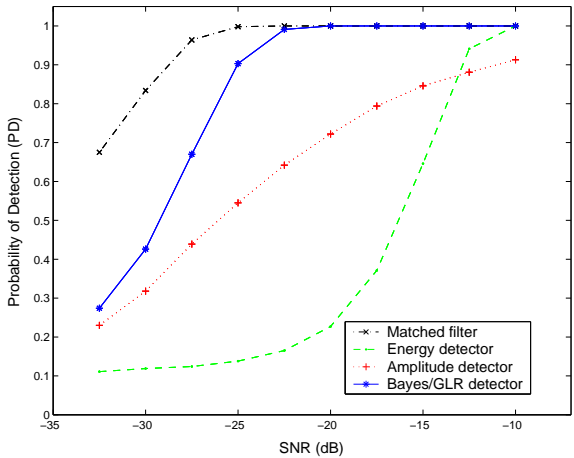}
%\end{minipage}
%
\caption{\small \textit{The power curves ($P_D$ vs.~SNR) for the four
    detectors studied in this paper for $P_F=0.1$ and $\lambda=1$
    event-per-second. At $P_D=0.8$ the hybrid Bayes/GLR detector performs
    within 4 dB of the bound established by the matched filter.}}
\label{fig:power}
\end{center}
\end{figure}

We ran the Gibbs sampler for 5,000 iterations for the hybrid 
Bayes/GLR detector. Fig.~\ref{fig:ROC15_2} and \ref{fig:ROC15_2p} 
show ROC curves for SNR = -25 dB and -20 dB, respectively, for 
$\lambda=1$ event-per-second. In both cases, our hybrid Bayes/GLR 
detector significantly outperformed all the other detectors 
except for the unimplementable matched filter.  The matched 
filter had complete information about the random flip times, and 
as a result it achieved almost perfect detection for both SNR 
values. In Fig.~\ref{fig:ROC175_2} the value of $\lambda$ was 
increased to 10 events-per-second and the SNR was held at -20 dB. 
As expected, the performance of the amplitude and hybrid 
Bayes/GLR detectors degrades, while the matched filter and energy 
detector, whose performance does not depend on $\lambda$, are not 
affected. In Fig.~\ref{fig:power} the power curves for all 
detectors are plotted as a function of SNR for $\lambda=1$.  Here 
all detectors perform at the same false alarm rate $P_F=0.1$, and 
we can make a quantitative SNR comparison by fixing the detection 
performance level at $P_D=0.8$, say. To attain this detection 
level, the energy detector and amplitude detector require SNR's 
of at least -14 dB and -17.5 dB, respectively, while the hybrid 
Bayes/GLR detector only requires -26 dB.  As compared to the 
amplitude detector, this represents an improvement of almost 9 dB 
in SNR performance using our proposed detector. Furthermore, the 
performance of our proposed detector is only 4 dB worse than the 
performance bound of -30 dB established by the matched filter for 
this level of $P_F$ and $P_D$.  Note that the amplitude detector 
outperforms the energy detector for low SNR's but not for high 
SNR's.This is explained by the fact that even though the energy 
detector is not affected by random flips, at low SNR's its test 
statistic is dominated by the noise.

\begin{figure}[!htb]
\begin{center}
%\begin{minipage}[b]{1.0\linewidth}
%\centering
%\includegraphics[width=3.0in]{Gibbsc.eps} 
\includegraphics[width=3.0in]{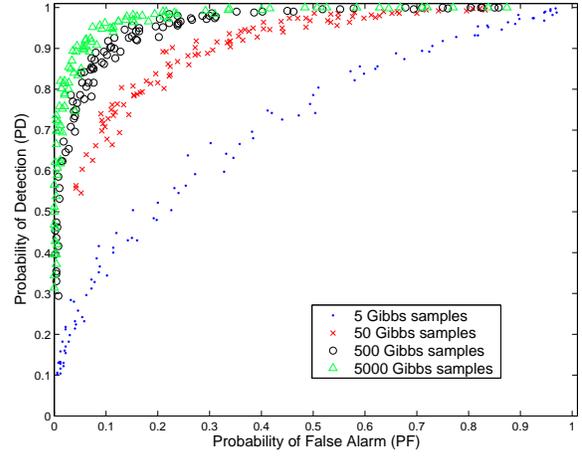}
%\end{minipage}
%
\caption{\small \textit{Juxtaposition of ROC curves of hybrid Bayes/GLR
detector, obtained with different numbers of Gibbs samples in the
maximization step, at SNR = -20 dB and $\lambda=10$
events-per-second. Performance improves as the number of Gibbs samples
increases.}}
\label{fig:GibbsN}
\end{center}
\end{figure}

In another simulation, we investigated the role of the number of Gibbs
samples on performance of the hybrid Bayes/GLR detector, shown in
%SNR was fixed at -17.5 dB. 
Fig.~\ref{fig:GibbsN}. It is evident that performance improves as we
increase the number of Gibbs samples. For example, at $P_{F}=0.1$, $P_{D}$
increases from approximately 0.35 to 0.65 if we increase the number of Gibbs
samples from 100 to 500. It increases further to around 0.9 and 0.95 if 500
or 5,000 Gibbs samples are used, respectively. Such improvements in
performance are significant but yield diminishing returns as the number of
Gibbs samples is increased beyond 500.


\section{Conclusion}

In this paper we presented a hybrid Bayes/GLR approach to 
detecting the presence of single spins for the OSCAR MRFM 
experiment. We have shown by simulation that the Bayes/GLR 
detector performs significantly better than the classical 
amplitude and energy detectors. The improvement in detection 
performance is due to the fact that, unlike the classical 
detectors, the new detector estimates the unknown values of the 
random spin reversal times and the initial polarity. Of course 
this performance improvement comes at the price increased 
implementation complexity. This complexity increases in the 
random reversal rate $\lambda$ due to the necessity to perform 
Gibbs sampling over an increasingly large number of probable spin 
reversal sequences.  Nonetheless, for the experiments in the first simulation with $\lambda=1$ event-per-second, the run time of our detector (5,000 Gibbs samples) was only 
on the order of about half a minute per 3-second measurement record (our code 
was implemented in Matlab 6.5, under WindowsXP on a 2.26GHz PC, with 510MB RAM).

An interesting extension of our results would be to assume that
the frequency shift $|\Delta\omega_o|$ is also unknown.  This would lead to a
hybrid Bayes/GLR detector which detects the peak over the spectrum of the
signal in addition to maximizing over the number and positions of the
transitions.

 The hybrid Bayes/GLR detector was derived using a baseband signal model
consisting of a random telegraph wave with Poisson transitions and AWGN. This
signal model is theoretically justified under the spin-lock assumption. The
validity of the spin-lock assumption remains to be established.  More
sophisticated signal models of the cantilever measurements, and associated
detection methods which bypass frequency demodulation and operate directly on
those measurements, are currently under investigation.

%\bibliography{/z/hero/TEXT/lib,lib} 
\bibliography{yip_paper6_3.bbl} 
\bibliographystyle{unsrt} 

\end{document}